# Using Developer Discussions to Guide Fixing Bugs in Software

**Sheena Panthaplackel**[1]**, Milos Gligoric**[2]**, Junyi Jessy Li**[3]**, Raymond J. Mooney**[1]

[1]Department of Computer Science
[2]Department of Electrical and Computer Engineering
[3]Department of Linguistics
The University of Texas at Austin
spantha@cs.utexas.edu, gligoric@utexas.edu
jessy@austin.utexas.edu, mooney@cs.utexas.edu

## Abstract

Automatically fixing software bugs is a challenging task. While recent work showed that natural language context is useful in guiding bug-fixing models, the approach required prompting developers to provide this context, which was simulated through commit messages written *after* the bug-fixing code changes were made. We instead propose using bug report discussions, which are available *before* the task is performed and are also naturally occurring, avoiding the need for any additional information from developers. For this, we augment standard bug-fixing datasets with bug report discussions. Using these newly compiled datasets, we demonstrate that various forms of natural language context derived from such discussions can aid bug-fixing, even leading to improved performance over using commit messages corresponding to the oracle bug-fixing commits.

## 1 Introduction

Software defects, or *bugs*, arise for a number of reasons, including missing or changing specifications, programming errors, poor documentation, and overall complexity (Rodríguez-Pérez et al., 2020). Due to the extensive developer time and effort needed to fix bugs (Weiss et al., 2007), there is growing interest in automated bug fixing (Tufano et al., 2019; Chen et al., 2019; Lutellier et al., 2020; Mashhadi and Hemmati, 2021; Allamanis et al., 2021; Chakraborty and Ray, 2021).

Most of these approaches only consider the buggy code snippet when generating the fix. However, with such limited context, this is extremely challenging. For instance, in Figure 1a, generating the fixed code requires removing .append("\n"), but this is not obvious from inspecting the buggy code alone. To address this, Chakraborty and Ray (2021) proposed prompting developers for a natural language description of intent (e.g., "Removed trailing newlines...") that can guide a model in performing the task. As a proxy, in their study,

```
void emptyImplicitTable(String table, int line) {
    sb.append("Invalid table definition due to
            empty implicit table name: ")
        .append(table)
        .append("\n");                          Buggy

    sb.append("Invalid table definition due to
            empty implicit table name: ")
        .append(table);                          Fixed
}
```
**Oracle Commit Message:** Removed trailing newlines from error messages. Fixes https://github.com/mwanji/toml4j/issues/18

(a) Buggy and fixed code snippets in `emptyImplicitTable` method with commit message for the oracle bug-fixing commit

**Title:** Parsing exception messages contain trailing newlines

**Utterance #1:**
Some of the parsing exceptions thrown by toml4j contains trailing newlines. This is somewhat unusual, and causes empty lines in log files when the exception messages are logged…

**Utterance #2:**
The idea was to be able to display multiple error messages at once. However, processing stops as soon as an error is encountered, so that's not even possible. Removing the newlines shouldn't be a problem, then.

**Solution Description:** remove trailing newlines from toml4j log messages

(b) Bug report discussion and generated solution description

Figure 1: Bug-fixing patch from the toml4j project, with context from the corresponding bug report discussion.

they used the commit message corresponding to the oracle commit which fixed the bug.

By showing that natural language can aid bug-fixing, their study yields promising results. However, we raise two concerns with their approach. First, prompting developers for additional information can be burdensome for them, as it requires time and manual effort. Second, and more importantly, it is unrealistic to use the oracle commit message as a proxy. Since it is written *after* the bug is fixed to document the code changes (Tao et al., 2021), it does not accurately reflect information actually available when the task needs to be performed.

In reality, there are more appropriate sources of natural language to guide fixing bugs, which are *naturally* occurring and available *before* the task is to be performed. Namely, many bugs are first reported through issue tracking systems (e.g., GitHub Issues), where developers engage in a *discussion* to collectively understand the problem, investigate the cause, and formulate a solution (before they are fixed) (Noyori et al., 2019; Arya et al., 2019).

Content in these discussions are often relevant to generating the fix. For example, in Figure 1b, the title suggests that the bug pertains to "trailing newlines" and the last utterance of the discussion recommends "removing the newlines." Additionally, using modern techniques (Panthaplackel et al., 2022) that summarize content relevant towards implementing the solution in a bug report discussion, we can also automatically obtain a natural language description of the solution ("remove trailing newlines..."). Note that these sequences provide insight on the intent of the fix, much like the oracle commit message, without requiring any additional input or any context beyond what is naturally available.

In this work, we use bug report discussions to facilitate automated bug fixing. While these discussions have been previously used to automate tasks related to bug resolution, such as localizing bugs (Koyuncu et al., 2019; Zhu et al., 2020) and assigning relevant developers (Xi et al., 2018; Baloch et al., 2020), to our knowledge, they have never been used to directly generate the bug-fixing code.

We propose various input context representations, encompassing different natural language components that are tied to the discussions and likely to capture their meaningful aspects. We *heuristically* derive components from discussions, including the discussion as a whole, the title, and last utterance. We also derive components *algorithmically* through model-generated solution descriptions and the attended discussion utterances during this generation. We incorporate these representations into large sequence-to-sequence models pretrained on large amounts of source code and technical text (Ahmad et al., 2021) and then do task-specific finetuning.

For training and evaluation, we mine bug report discussions from GitHub Issues and map them to subsets of Tufano et al. (2019)'s bug-fixing patches datasets.[1] Results show that when bug report discussions are available, they lead to significant improvements in fixing bugs, even outperforming using the oracle commit message.

## 2 Using Bug Report Discussions

Many bugs are reported with issue tracking systems, through which a user can open a bug report and initiate a discussion with developers. The user first states the problem in the *title* and typically elaborates in the first *utterance*. Developers then join the discussion and engage in a dialogue with the user as well as other developers.

These discussions isolate the problem, diagnose the cause, and prescribe potential solutions (Arya et al., 2019). Due to their technical nature, they often span more than just natural language, including system error messages and relevant code snippets (Li et al., 2018). Furthermore, they are readily available *before* bugs are fixed. So, we consider using these contextually rich discussions to guide the task of bug fixing. We devise various strategies for heuristically and algorithmically deriving context from these discussions.

### 2.1 *Heuristically* Deriving Context

We consider using the *whole discussion*, including the title and all utterances (occurring before the bug-fixing code changes are implemented). However, these discussions can be extremely long (Table 1), making them difficult for neural models to reason about and also extending beyond the input length capacities of many models (e.g., 1,024 tokens) (Ahmad et al., 2021) in some cases. For this reason, we look at more concise elements within the discussion which might convey its meaningful aspects. First, we consider the *title*, as it is a brief summary of the bug (Chen et al., 2020). Next, we consider the *last utterance* before the bug-fixing commit, since it captures the most recent information and also roughly corresponds to the point at which a developer acquired enough context about the fix to implement it (Panthaplackel et al., 2022).

### 2.2 *Algorithmically* Deriving Context

To guide developers in absorbing information relevant towards implementing the solution for a given bug report, we recently proposed generating a brief natural language description of the solution by synthesizing relevant content from within the whole bug report discussion (Panthaplackel et al., 2022).

To generate these *solution descriptions*, we finetuned a large pretrained encoder-decoder model. For training supervision of solution descriptions, we used commit messages and pull request titles corresponding to the commits and pull requests linked to bug reports. To control for noise, we relied on a filtered training set, consisting of fewer generic and uninformative target descriptions as well as discussions without sufficient context to generate informative descriptions). We provide

---
[1] Data is available at `https://github.com/panthap2/developer-discussions-for-bug-fixing`.

additional details of our approach for generating solution descriptions in Appendix A.

While these solution descriptions are intended to guide humans in manually fixing bugs, we evaluate whether they can also guide models in automatically performing the task. Furthermore, since the title corresponding to the bug report discussion and the solution description summarize different aspects of the discussion, we investigate the benefits of combining the two (solution description + title).

Next, the segments (title or individual utterances) from the discussion that contribute the most towards generating a natural language description of the solution are likely to also be useful towards implementing that solution (i.e., generating the fix). To approximate the most relevant discussion segments, we use attention. Namely, we examine the last layer of Panthaplackel et al. (2022)'s decoder to determine the most highly attended input token at each decoding step and the segment (title or individual utterance) to which it belongs. From this, we obtain the *attended segments*.

## 3 Data

Chakraborty and Ray (2021) relied on the commonly used bug-fixing patches (*BFP*) datasets (Tufano et al., 2019). This entails $BFP_{small}$, with examples extracted from Java methods spanning fewer than 50 tokens, and $BFP_{medium}$, with examples extracted from methods spanning 50-100 tokens. In this work, we also focus on these datasets, particularly the preprocessed versions released by Chakraborty and Ray (2021). However, since they do not include the associated bug report discussions, we enrich examples with this information.

### 3.1 Mining Bug Report Discussions

We mine issue reports from GitHub Issues, for the 58,597 projects that encompass examples in the *BFP* datasets. We obtain 1,878,096 issue reports, 365,005 of which are linked to commits made between March 2011 and October 2017 (time frame used for mining the *BFP* datasets). By matching these commits to the *bug-fixing* commits from which the *BFP* examples were drawn, we identify the examples that correspond to *bug* reports. We map 3,028 (of the 58,287) examples in $BFP_{small}$ and 3,333 (of the 65,404) examples in $BFP_{medium}$ to bug report discussions, forming the *discussion-augmented bug-fixing patches* (*Disc-BFP*) datasets: $Disc\text{-}BFP_{small}$ and $Disc\text{-}BFP_{medium}$.

|  | $Disc\text{-}BFP_{small}$ | $Disc\text{-}BFP_{med}$ |
| --- | --- | --- |
| #Ex | 3,028 | 3,333 |
| #Discussions/Ex | 1.3 | 1.3 |
| #Utterance/Discussion | 2.8 | 2.9 |
| #Attn Segments/Ex | 1.0 | 1.0 |
| Buggy | 22.1 | 42.4 |
| Fixed | 19.3 | 40.8 |
| Method | 32.2 | 74.2 |
| Oracle Msg | 19.7 | 19.6 |
| Title | 7.9 | 8.1 |
| Utterance | 127.6 | 136.4 |
| Last Utterance | 114.0 | 109.3 |
| Soln Desc | 8.5 | 8.5 |

Table 1: ***Disc-BFP* dataset statistics**. We report averages across all data splits. Average token lengths (split by punctuation and spacing) are presented in the second block. Note that we consider only utterances occurring before the bug-fixing commit.

Note that *Disc-BFP* is comparatively smaller than *BFP*. While constructing *BFP*, Tufano et al. (2019) did not consider any mining criteria related to bug reports, so it is not surprising that many of their examples do not have bug report discussions. Bugs can be identified through various development activities like code review, testing, and bug reporting. In this work, we focus on the last scenario, for which bug report discussions would naturally be available.

### 3.2 Data Processing

$Disc\text{-}BFP_{small}$ consists of 2,445 training, 290 validation, and 293 test examples. $Disc\text{-}BFP_{medium}$ consists of 2,660 training, 341 validation, and 332 test examples. In doing this, we maintain the original data splits (e.g., $Disc\text{-}BFP_{small}$'s training set is strictly a subset of $BFP_{small}$'s training set).

A bug report discussion is organized as a timeline, and we consider only content that precedes the bug-fixing commit on the timeline, corresponding to the naturally-available context. Since a commit can be linked to multiple issue reports, some examples have multiple bug report discussions. In these cases, we order them so that discussions with the most recent activity appear first and are less likely to get truncated due to input length constraints (as explained in the next paragraph). When leveraging individual discussion components (e.g., title, generated solution description), we derive them from each discussion separately and concatenate them (separated with <s>). We process bug report bug report discussions similar to Panthaplackel et al. (2022), and we use the processed *BFP* code data (buggy code and method) released by Chakraborty and Ray (2021). We present dataset statistics in Table 1.

## 4 Models

Chakraborty and Ray (2021) achieved state-of-the-art performance on the *BFP* datasets by finetuning PLBART (Ahmad et al., 2021), a large sequence-to-sequence model that was pretrained as a denoising autoencoder (Lewis et al., 2020) on large amounts of source code from GitHub and technical text from StackOverflow. Similarly, we consider finetuning PLBART to generate the fixed code given varying input context representations.

### 4.1 Model Initialization

Since Chakraborty and Ray (2021) finetuned using significantly more data (i.e., *BFP* training sets)[2], we initialize models using their checkpoints that were finetuned with the buggy code snippet and the full method context (`emptyImplicitTable` in Figure 1a): *buggy* <s> *method*. This helps contextualize the buggy code snippet and was shown to improve performance.[3]

### 4.2 Our Models

After initializing, we further finetune on the *Disc-BFP$_{small}$* and *Disc-BFP$_{medium}$* training sets (separately). All input context representations used for this are formed by concatenating *buggy* <s> *method* <s> with the various natural language sequences tied to bug report discussions outlined in Section 2. Sequences entailing multiple elements (e.g., utterances in the whole discussion, titles from multiple bug report discussions) are separated with <s>.

Though PLBART is capable of handling up to 1,024 tokens as input, Chakraborty and Ray (2021) limit to 512 tokens. However, since the sequences we consider can be particularly long after the SentencePiece tokenization (Kudo and Richardson, 2018) employed by PLBART, we choose to utilize the full capacity during our finetuning. Note that the input is truncated by removing from the end if it exceeds the limit. We provide additional details regarding model training in Appendix E.

### 4.3 Baselines

We consider models which use only *buggy* <s> *method* (without natural language). As points of reference, we also consider models that use the oracle commit message rather than context from bug report discussions: *buggy* <s> *method* <s> *oracle commit message*. To make a fair comparison with our models, we initialize baselines using the Chakraborty and Ray (2021) checkpoints (§4.1) and further finetune on the *Disc-BFP* training sets, using a context window of 1,024 tokens.

---

[2]We considered directly finetuning PLBART on the smaller *Disc-BFP* training sets, but this resulted in relatively low performance, as shown in Appendix D.

[3]Note that the method also contains the buggy code snippet. Though repetitive, this outperformed a unified format.

| Finetune/Test Context | *Disc-BFP$_{small}$* | *Disc-BFP$_{med}$* |
|---|---|---|
| Without NL* | 33.8 | 27.1 |
| Oracle Msg† | 33.4 | 27.4 |
| Whole Discussion | 33.1 | 27.1 |
| Title | 35.5*† | 25.9 |
| Last Utterance | 35.2*† | **28.9*†** |
| Soln Desc | 33.8 | 27.4 |
| Soln Desc + Title | 35.5*† | 25.6 |
| Attended Seg | **36.2*†** | 28.0* |

Table 2: Results on the *Disc-BFP* test sets. Models are initialized from the checkpoint originally finetuned **without the oracle commit message** on the full *BFP* training sets. We then finetune on the *Disc-BFP* training sets with various input context representations and evaluate on the *Disc-BFP* test sets using the same representations. We indicate representations that statistically significantly outperform baselines with superscripts identifying the specific baseline that is surpassed.

## 5 Results

Following Chakraborty and Ray (2021), we compute how often (%) the generated output *exactly matches* the target fixed code snippet. We perform statistical significance testing with bootstrap tests (Berg-Kirkpatrick et al., 2012), using 10,000 samples (with sample size 5,000) and $p < 0.05$. We provide sample output in Appendix B.

We present results in Table 2. We find that leveraging context from bug report discussions can lead to significant improvements over baselines which do not include natural language context, yielding up to 2.4% improvement for *Disc-BFP$_{small}$* and 1.8% for *Disc-BFP$_{medium}$*.

We also observe that using context derived from bug report discussions leads to improved performance (1.5-2.8%) over using the oracle commit message during our finetuning with the *Disc-BFP* training sets. However, when Chakraborty and Ray (2021) originally considered using the oracle commit message, they had finetuned with it as input on significantly more data (i.e., the full *BFP* training sets). So, we further investigate by initializing PLBART parameters from the Chakraborty and Ray (2021) checkpoint which was finetuned using the oracle commit message (*buggy* <s> *method* <s> *oracle commit message*). Then, we perform finetuning on the *Disc-BFP* training sets with the various input context representations we considered in Table 2. We present results from these additional experiments in Table 3. Note that for all representations other than "oracle msg" the oracle commit message is used only during training and not used at test time, and so the results can extend to an actual realistic use case.

| Finetune/Test Ctxt | $Disc\text{-}BFP_{small}$ | $Disc\text{-}BFP_{med}$ |
|---|---|---|
| Without NL[§] | 35.5 | 25.3 |
| Oracle Msg[¶] | 36.2 | 25.9 |
| Whole Discussion | 34.1 | 25.6 |
| Title | 35.2 | 25.3 |
| Last Utterance | 36.2 | 25.6 |
| Soln Desc | 33.4 | **26.5**[§] |
| Soln Desc + Title | **39.2**[§¶] | 26.2[§] |
| Attended Seg | 36.9[§] | 24.1 |

Table 3: Additional results on the *Disc-BFP* test sets where models are initialized from the checkpoint originally finetuned **with the oracle commit message** on the full *BFP* training sets. We then finetune on the *Disc-BFP* training sets with various input context representations and evaluate on the *Disc-BFP* test sets using the same representations. We indicate representations that statistically significantly outperform baselines with superscripts identifying the specific baseline that is surpassed.

Relative to the results presented in Table 2, for *Disc-BFP$_{small}$*, initializing from the checkpoint finetuned with the oracle commit message tends to yield improved performance across the different input context representations, and with the solution description + title representation, we observe a 3.0% improvement over using the oracle commit message. For *Disc-BFP$_{medium}$*, the performance tends to be lower, and the "last utterance" context representation from Table 2 remains the best. Therefore, we again find that using bug report discussions leads to improvements over baselines that use the oracle commit message (during both finetuning and test). This suggests that context derived from bug report discussions, encompassing diverse types of information, can offer richer context than oracle commit messages for fixing bugs. This is especially promising since these discussions are often readily available in a real world setting.

Overall, the scores and magnitude of improvement tend to be lower for *Disc-BFP$_{medium}$*. This is likely due to the challenges of generating longer sequences (Varis and Bojar, 2021) and the stringent evaluation metric requiring exact match with the reference. The best performance on the *Disc-BFP$_{small}$* test set comes from using solution description + title. For *Disc-BFP$_{medium}$*, it is with the last utterance. Since both of these are derived from the whole discussion, one may expect using the whole discussion to yield similar or even improved performance; however, this is not the case.

Including the whole discussion substantially increases the input length, which models like PLBART cannot easily handle. This can be partially attributed to the practical challenge of fitting the entire sequence in the model's limited context window, with 12.8-15.8% training examples getting truncated. However, the bigger challenge is drawing meaning from such large amounts of text. We demonstrate the benefits of using more concise sequences, through various natural language elements that are likely to capture critical aspects of the whole discussion.

## 6 Conclusion

In this work, we investigated the utility of natural language for automated bug fixing. Unlike prior work, which leverages an unrealistic source of natural language for this purpose, through oracle commit messages, we consider a naturally occurring source that is often available: bug report discussions. We explore various strategies for deriving natural language context from these discussions, using our newly compiled discussion-augmented, bug-fixing patches datasets. We show that when these discussions are available, they offer useful context for bug fixing, even leading to improved performance over using oracle commit messages.

## Acknowledgements

We would like to thank Saikat Chakraborty for giving us access to checkpoints from Chakraborty and Ray (2021). We would like to also thank anonymous reviewers for their detailed suggestions. This work was supported by NSF grant IIS-2145479, a Bloomberg Data Science Fellowship to the first author, and a Google Faculty Research Award.

## Limitations

We focus on popular bug-fixing datasets (Tufano et al., 2019), which were originally constructed with certain constraints, including the use of a single programming language (Java) and methods of limited lengths (<50 tokens, 50-100 tokens). Next, examples in these datasets correspond to individual methods, with some examples being drawn from different methods of the same bug-fixing commit. Therefore, while generating the correct fix for a given example removes the presence of a bug in a particular method, it does not necessarily imply that the underlying bug has been completely removed from the software project entirely.

Furthermore, bug report discussions are *not always* available, and our work focuses on those instances in which they are available. Because Tufano et al. (2019) do not consider bug report discussions in their work, they do not require examples to have bug report discussions in their datasets. However, we do need examples to have these discussions for our study. Because we are unable to map many of their examples to bug report discussions, we focus on smaller subsets of their datasets.

## Ethics Statement

Automated bug fixing aims to streamline debugging and bug resolution for developers. We envision developers using the output generated by our models as "suggested fixes" that they would still need to inspect (and possibly revise) before committing them to the code base. Without such human intervention, erroneous output generated by our models could leave bugs unfixed or even introduce new bugs, posing a threat to the overall reliability of the software.

Note that we mine publicly available bug report discussions, in accordance with GitHub's acceptable use policy.

## A  Generating Solution Descriptions

In Panthaplackel et al. (2022), we benchmarked various models, finding that the best solution descriptions were generated by finetuning PLBART with a filtered training set (consisting of fewer generic and uninformative target descriptions as well as discussions without sufficient context to generate informative descriptions).

In this current work, we re-train the model after removing 7 examples in the training set that have bug reports overlapping with the *Disc-BFP* test sets. We run inference on all partitions of the *Disc-BFP* datasets. For this, we first preprocess the bug report discussions by *subtokenizing* them (i.e., splitting by spaces, punctuation, camelCase, and snake_case), similar to how we previously preprocessed the training data in Panthaplackel et al. (2022). Note that we do not subtokenize bug report discussions when we directly feed them into the models we presented in the main paper. Bug report discussions often include source code, either in-lined with natural language or as longer code blocks, which are often delimited with markdown tags. In Panthaplackel et al. (2022), we had retained in-lined code but removed longer marked blocks of code. While these longer code blocks may not be as relevant to generating natural language descriptions, we believe they could be useful in gathering insight for generating the fixed code. Therefore, we do not remove them from bug report discussions, even when generating solution descriptions.

## B  Examples

For the *Disc-BFP$_{small}$* test example in Figure 1, the two models which leverage only *buggy* <s> *method* during finetuning and test (Without NL in Table 2) do not generate the correct output. Note that neither of these models have access to any natural language context. Two other models (which do use natural language) also fail to generate the correct output, corresponding to the whole discussion and solution description representations (initialized using the "Without NL" checkpoint). In all four of these error cases, the model simply copies the buggy code snippet. However, the other 12 models generate the correct output for this particular example.

Some examples are difficult for models, even with natural language context. We provide one such example from the *Disc-BFP$_{medium}$* test set in Figure 2. The fix requires reversing the order of the method parameters, which is actually evident from the bug report discussion, as well as the generated solution description. However, performing this reversal involves more complex reasoning, and so the majority of models are unable to generate the correct output for this example. Nonetheless, the model which leverages the last utterance (initialized using the "Without NL" checkpoint) does manage to generate the correct output.

## C  Identifying Useful Discussion Segments

We acquire context from bug report discussions in various ways, either *heuristically* (whole discussion, title, last utterance) or *algorithmically* (attended segments when generating solution descriptions). (Note that we do not include solution descriptions in these groups since they do not actually appear within the bug report discussions.) As we saw in Table 2, using the whole discussion may not be beneficial, since models struggle to reason about large amounts of text. We show that we are able to achieve improved performance by selecting more concise segments from within this discussion (e.g., title, last utterance, attended segments) that are likely to be relevant to fixing the bug.

However, we may not always being selecting the most useful segment(s) yielding the best performance. The most useful segments may vary by example, and there could also be other utterances (beyond the title, last utterance, and attended utterances) that have relevant information.

Therefore, we also estimate the performance of an "oracle" upper-bound that employs the most useful segment as the natural language context. For this, we consider models finetuned with the various segments from the discussion, including models finetuned on the whole discussion. We run inference with these models, using *buggy* <s> *method* <s> *segment*, for all segments, including the title and each utterance in the discussion. So, if there are $N$ segments derived from the discussion (title and $N-1$ utterances before the bug-fixing commit), we obtain $N$ candidates for the fixed code.

For a given example, we compute *best exact match*, or how often *at least one* of these candidates matches the reference. We present results in Table 4. We observe a 3.1–3.3% gap, relative to the highest scores in Table 2, suggesting that there is useful context in these discussions that is not being exploited. We leave it to future work to learn models for extracting the most useful segments from bug report discussions for fixing bugs.

```
- public void assertEquals (java.lang.Object actual, java.lang.Object expected)
+ public void assertEquals (java.lang.Object expected, java.lang.Object actual)
{
    if ((expected == null) && (actual == null))
        return;
    if ((expected != null) && (expected.equals(actual)))
        return;
    fail(format(expected, actual));
}
```

**Oracle Commit Message:** Fixes issue #4.

**Title:** assertEquals parameters order

> **Utterance #1:**
> Maybe this is not an issue but a desired behaviour, however, it seems to me that the order of the parameters in the assertEquals method is wrong: public void assertEquals(Object actual, Object expected) Being a long time user of JUnit, I expected the "actual" parameter to be in the second position instead of the first one.

> **Utterance #2:**
> The ordering was based on TestNG, which is what I typically use for unit testing, but since xUnit is more common I don't mind reversing the order.

**Solution Description:** reversing the order of the assert equals parameters

Figure 2: Examples from the *Disc-BFP$_{medium}$* test set, with the corresponding bug report discussion (https://github.com/jhalterman/concurrentunit/issues/4) and generated solution description.

| Init | Finetune Context | *Disc-BFP$_{small}$* | *Disc-BFP$_{med}$* |
|---|---|---|---|
| Without NL (*BFP*) | Whole Discussion | 36.9 | 29.8 |
| | Title | 40.3 | 27.4 |
| | Last Utterance | 36.9 | **32.2** |
| | Attended Segments | 37.2 | 31.3 |
| With NL (*BFP*) | Whole Discussion | 39.6 | 29.2 |
| | Title | 38.2 | 26.8 |
| | Last Utterance | 39.2 | 29.5 |
| | Attended Segments | **42.3** | 27.1 |

Table 4: Evaluating exact match (%) if the best performing segment (title or any individual utterance) from the whole discussion is used at test time (assuming that it's known).

## D Initializing Model Parameters

In Table 2, we present results from initializing model parameters from two of the checkpoints released by Chakraborty and Ray (2021). One corresponds to finetuning PLBART *without NL* using task-specific data from the larger *BFP* training sets. The other one corresponds to finetuning PLBART *with NL* (from oracle commit messages), also using task-specific data from the *BFP* training sets. Since these checkpoints have already been finetuned on bug-fixing data, it is reasonable to run inference on them directly without further finetuning on the *Disc-BFP* training sets. We show these results in Table 5. We find the overall performances to be lower, especially when testing with input context representations that were not seen during Chakraborty and Ray (2021)'s finetuning (e.g., whole discussion).

We also tried initializing model parameters directly from PLBART and finetuning on the *Disc-BFP* training sets. Table 5 shows that this works poorly, likely because the *Disc-BFP* training sets are smaller than the *BFP* training sets, with which the Chakraborty and Ray (2021) checkpoints were finetuned. Therefore, to reap the benefits of finetuning on more data, we believe it is best to first finetune on larger bug-fixing datasets (for which bug report discussions do not need to be available). Following that, another stage of finetuning should be done using the smaller training set that includes context from bug report discussions.

## E Training Details

Our models are based on the architecture of PLBART, which itself follows from the BART-base model (Lewis et al., 2020). The encoder and decoder each have 6 layers, with hidden dimension 768 and 12 heads. There are approximately 140M parameters. We use the same hyperparameters as Chakraborty and Ray (2021). The batch size is 4, with gradient accumulation over every 4 batches. Early stopping is employed, with a patience of 5 epochs, based on validation performance. All models are trained for a single run. At test time, beam search is used, with a beam size of 5. Models are finetuned and tested using NVIDIA DGX GPUs (32 GB). We report the number of epochs, training time, and testing time for each of the models in Table 6.

| Init | Context | Inference Only | | Finetuned | |
|---|---|---|---|---|---|
| | | $Disc\text{-}BFP_{small}$ | $Disc\text{-}BFP_{med}$ | $Disc\text{-}BFP_{small}$ | $Disc\text{-}BFP_{med}$ |
| PLBART | Without NL | - | - | 22.2 | 14.8 |
| | Oracle Msg | - | - | 28.0 | 16.6 |
| | Whole Disc | - | - | 25.3 | 16.0 |
| | Title | - | - | 27.3 | 1.5 |
| | Last Utterance | - | - | 23.2 | 19.0 |
| | Soln Desc | - | - | 20.5 | 17.2 |
| | Soln Desc + Title | - | - | 24.2 | 16.3 |
| | Attended Seg | - | - | 18.1 | 1.8 |
| Without NL (*BFP*) | Without NL | 30.7 | 25.3 | 33.8 | 27.1 |
| | Oracle Msg | 30.7 | 25.0 | 33.4 | 27.4 |
| | Whole Disc | 21.5 | 19.0 | 33.1 | 27.1 |
| | Title | 31.4 | 25.9 | 35.5 | 25.9 |
| | Last Utterance | 29.0 | 23.2 | 35.2 | **28.9** |
| | Soln Desc | 31.7 | 25.9 | 33.8 | 27.4 |
| | Soln Desc + Title | 29.7 | 25.0 | 35.5 | 25.6 |
| | Attended Seg | 23.5 | 20.8 | 36.2 | 28.0 |
| With NL (*BFP*) | Without NL | 31.1 | 22.3 | 35.5 | 25.3 |
| | Oracle Msg | 31.1 | 24.4 | 36.2 | 25.9 |
| | Whole Disc | 20.5 | 16.9 | 34.1 | 25.6 |
| | Title | 28.7 | 22.3 | 35.2 | 25.3 |
| | Last Utterance | 25.3 | 22.3 | 36.2 | 25.6 |
| | Soln Desc | 29.4 | 24.1 | 33.4 | 26.5 |
| | Soln Desc + Title | 28.3 | 22.6 | **39.2** | 26.2 |
| | Attended Seg | 23.5 | 19.9 | 36.9 | 24.1 |

Table 5: We measure the effect of finetuning on the *Disc-BFP* training sets by comparing to a setting in which the Chakraborty and Ray (2021) checkpoints are used directly for inference (without any finetuning). We also measure the effect of initializing with checkpoints that have already been finetuned on task-specific data by comparing to models directly initialized from PLBART and then finetuned on the *Disc-BFP* training sets.

| Init | Context | $Disc\text{-}BFP_{small}$ | | | $Disc\text{-}BFP_{med}$ | | |
|---|---|---|---|---|---|---|---|
| | | **Epoch** | **Train Time** | **Test Time** | **Epoch** | **Train Time** | **Test Time** |
| Without NL (*BFP*) | Without NL | 2 | 0:15:44 | 0:01:06 | 3 | 0:34:43 | 0:01:59 |
| | Oracle Msg | 14 | 0:31:11 | 0:00:34 | 4 | 0:28:21 | 0:01:13 |
| | Whole Disc | 6 | 0:25:25 | 0:01:55 | 10 | 1:02:34 | 0:02:18 |
| | Title | 6 | 0:18:54 | 0:01:31 | 13 | 1:05:32 | 0:01:02 |
| | Last Utterance | 8 | 0:28:28 | 0:01:38 | 5 | 0:37:44 | 0:02:47 |
| | Soln Desc | 4 | 0:17:48 | 0:01:11 | 3 | 0:26:54 | 0:01:50 |
| | Soln Desc + Title | 10 | 0:28:18 | 0:01:16 | 9 | 0:51:09 | 0:02:27 |
| | Attended Seg | 6 | 0:22:54 | 0:01:21 | 2 | 0:27:37 | 0:02:39 |
| With NL (*BFP*) | Without NL | 6 | 0:17:54 | 0:00:53 | 3 | 0:24:16 | 0:01:18 |
| | Oracle Msg | 9 | 0:36:17 | 0:00:52 | 9 | 1:02:02 | 0:01:59 |
| | Whole Disc | 6 | 0:25:55 | 0:00:53 | 5 | 0:42:01 | 0:02:19 |
| | Title | 6 | 0:18:08 | 0:01:31 | 10 | 0:46:07 | 0:01:36 |
| | Last Utterance | 4 | 0:19:51 | 0:01:54 | 2 | 0:24:24 | 0:02:09 |
| | Soln Desc | 2 | 0:13:44 | 0:01:03 | 2 | 0:23:42 | 0:01:50 |
| | Soln Desc + Title | 8 | 0:26:41 | 0:01:42 | 6 | 0:41:55 | 0:01:38 |
| | Attended Seg | 10 | 0:35:53 | 0:01:43 | 5 | 0:41:44 | 0:02:28 |

Table 6: Reporting the training epoch from which we obtain the checkpoint used for evaluation, the total training time (HH:MM:SS), and the total time needed to run inference (HH:MM:SS).